\documentclass[graybox,natbib,nosecnum]{svmult}
\bibpunct{(}{)}{;}{a}{}{,} % suppress commas between author-names and year

\pdfoutput=1   %forces use of pdflatex. Disable if you prefer to use .eps or .ps figures.
\usepackage{mathptmx}       % selects Times Roman as basic font
\usepackage{helvet}         % selects Helvetica as sans-serif font
\usepackage{courier}        % selects Courier as typewriter font
\usepackage{type1cm}        % activate if the above 3 fonts are
\usepackage{makeidx}         % allows index generation
\usepackage{graphicx}        % standard LaTeX graphics tool
\usepackage{multicol}        % used for the two-column index
\usepackage[bottom]{footmisc}% places footnotes at page bottom
\usepackage[normalem]{ulem}	% for strike-through of text with \sout{}  
\usepackage{hyperref}  %for hyperlinks

\usepackage{soul}   % for high-lighting of text
\newcommand{\hbindex}[1]{\hl{#1}\index{#1}}  %highlights index entries

\makeindex             % used for the subject index
\begin{document}

\title*{TRAPPIST-1 and its compact system of temperate rocky planets}
\author{Micha\"el Gillon}
\institute{Micha\"el Gillon \at Astrobiology Research Unit, University of Li\`ege, All\'ee du 6 ao\^ut 19, 4000 Li\`ege, Belgium. \email{michael.gillon@uliege.be}
}
\maketitle

\abstract{The TRAPPIST-1  system is comprised of seven Earth-sized rocky planets in small orbits around a Jupiter-sized ultracool dwarf star 12 parsec away. These planets cover an irradiation range  similar to the range of the inner solar system. Three of them orbit within the circumstellar habitable zone. All of them are particularly well-suited for detailed characterization, thanks to the small size of and to the infrared brightness of the host star, and to the system's compact resonant structure. An intense transit-timing monitoring campaign resulted in unprecedented precisions on the planets' masses and densities, and in strong constraints on their compositions. Transit transmission spectroscopy with HST discarded the presence of extended primary atmospheres around the seven planets. The first thermal emission measurements obtained with JWST favor \hbindex{low-density-atmosphere} or bare-rock scenarios for the two inner planets. The detection of dense secondary atmospheres around the five outer planets could be achieved by transit transmission spectroscopy with JWST, but this will require addressing the critical problem of stellar contamination with more theoretical and observational work.  }

\section{Introduction }

%Since the seminal discovery of 51 Pegasi b  \citep{MayorQueloz1995}, several thousands exoplanets have been detected at an ever increasing rate \citep{Schneider2011, Akeson2013}. While most efforts focused initially on the detection of new exoplanets, the field shifted gradually to the detailed characterization of specific planets, resulting notably in the first observational constraints on the atmospheric properties of extrasolar worlds. Most of these exoplanets detections and characterizations have relied on the \hbindex{transiting} configuration  of the detected/studies planets \cite{Winn2010}.
Atmospheric studies of \hbindex{transiting exoplanets} have so far been mostly limited to highly-irradiated gas giants \citep[see e.g.][]{Madhu2019} due to the instrumental limits of our observational facilities,  even if some pioneering results could already be gathered with the space telescopes {\it Spitzer} and {\it Hubble} (HST) for smaller and/or more temperate `super-Earths' and `mini-Neptunes' \citep[e.g.][]{Demory2012, Demory2016, Kreidberg2014, Kreidberg2019, Benneke2019}. But this situation is about to change.
Launched at the end of 2021 and fully operational since summer 2022, the \hbindex{JWST} space telescope has much better infrared coverage, sensitivity and resolution than HST and {\it Spitzer} that should enable dramatically progress in the study of transiting exoplanets, as demonstrated by its first published results \cite[e.g.][]{Ahrer2023, Alderson2023, Feinstein2023, Rustamkulov2023}. These exquisite instrumental capacities should even make it possible to extend such detailed characterization to temperate exoplanets similar in size to the Earth, provided that they transit very nearby (up to 20pc) and very-late-type (M5 or latter) M-dwarfs \citep{Kaltenegger2009, Charbonneau2009, dewit2013}. 

This `M-dwarf Opportunity' led to the development of several transit searches focused on nearby M-dwarfs like MEarth \citep{Nutzman2008} and TESS \citep{Ricker2016}. One of these projects, \hbindex{SPECULOOS} \citep{Gillon2018}, decided to focus on nearby `ultracool' dwarf stars (UDS) \citep{Kirkpatrick1997}, i.e. stars at the bottom of the main-sequence with spectral types later than M6, effective temperatures $\le$2800K, masses between 7 and 10\% the one of the Sun, and sizes similar to Jupiter. In 2015, the SPECULOOS prototype on the TRAPPIST telescope \citep{Gillon2013} detected two transiting Earth-sized planets in very-short orbits around \hbindex{TRAPPIST-1}, an M8-type UDS located about 12 parsec away \cite{Gillon2016}. Further observations from the ground and with the { \it Spitzer} and {\it Kepler} space telescopes revealed the existence of five more Earth-sized planets in the system \citep{Gillon2017, Luger2017}.

 TRAPPIST-1 is exceptional on many levels. It is indeed the system with the largest number of Earth-sized planets, and the only UDS planetary system,  known so far. It is extremely compact: its inner and outer planets orbit respectively at $\sim 0.01$ and $\sim 0.06$ au from their star, and the orbital distances between adjacent planets are only a few Earth-moon distances. Despite this extreme compactness, the stability of the system is sustained by a complex chain of mean-motion resonances \citep{Luger2017}. Its planets are the terrestrial exoplanets with the most precise mass and density measurements, resulting in tight constraints on their compositions \citep{Agol2021}. Three of these planets orbit within the habitable zone \citep{Kasting1993} of the star, and all of them are in theory amenable for the detection and study of a compact secondary atmosphere with JWST \citep[e.g.][]{Morley2017, Lustig-Yaeger2019}. 

The next sections summarize our current knowledge of the star TRAPPIST-1 and of its seven planets. 

\section{The ultracool dwarf star TRAPPIST-1}

TRAPPIST-1 = 2MASS J23062928-0502285 was discovered in 2000 by a search for nearby \hbindex{ultracool dwarfs} based on photometric selection criteria \citep{Gizis2000}, and identified as a high proper-motion and moderately  active $\sim$ M7.5 dwarf at about 11 parsecs from Earth. Subsequent studies converged on a spectral type of
M8.0 $\pm$ 0.5 \citep{Liebert2006, Bartlet2007, Gillon2016} and on a distance of 12.467 $\pm$ 0.011 pc \citep{GaiaEDR3}, while revealing a moderate level of activity typical of nearby very-late-type M-dwarfs \citep{Schmidt2007, Reiners2010, Lee2010}.

Following the discovery of its planetary system, many studies further constrained the star’s physical properties, including average density (from the planetary transits), bolometric luminosity, mass, radius, and effective temperature, while ruling out the
existence of wide stellar or brown dwarf companions \citep{ Howell2016, Gillon2017,  Luger2017, Paudel2018, Gonzales2019, VanGrootel2018A}. Analyses of the high-energy output of TRAPPIST-1 confirmed a significant quiescent emission in H$_\alpha$ ($\log_{10}(L_{\alpha}/L_{bol})$  between -4.85 and -4.70)  \citep{Gizis2000, Reiners2007, Barnes2014, Burgasser2015} and X-ray ($\log_{10}(L_{X}/L_{bol}) \sim -3.55$) \citep{Weathley2017}, while the star's surface magnetic field was estimated to be $600^{+200}_{-400}$ G by \cite{Reiners2007}. Subsequent observations in Lyman-$\alpha$ by HST \citep{Bourier2017a, Bourrier2017b} enabled an estimate of the ratio of the XUV and bolometric luminosities, $\log_{10}(L_{XUV}/L_{bol})$, to be around -3.5. It was also shown that the star's emission is significantly variable in the XUV and optical parts of the spectrum. This variability is dominated by frequent flares \citep{Gillon2017, Vida2017, Paudel2018}, with a flare of energy above $10^{29}$ erg taking place every 1-2 days, and a 'super-flare' of energy above $10^{33}$ erg occurring once every $\sim$ 3 months \citep{Paudel2018}. 

On top of the flares, optical time-series photometry of the star showed significant low-frequency variability. Using ground-based time-series photometry gathered by the TRAPPIST telescope in 2015, \cite{Gillon2016} first proposed P=1.40$\pm$0.05d as the rotation period of the star. Subsequently, the K2 space mission observed the star continuously for about 3 months, and Fourier analyses of the resulting photometry revealed a significant power excess at $\sim$3.3d, the 1d-alias of the value previously identified from ground-based data \citep{Luger2017, Vida2017}. Based on the fact that this modulation was not detected in the $>$1000hrs light curve of the star obtained at 4.5 $\mu$m by Spitzer \citep{Gillon2017, Delrez2018, Ducrot2020}, \cite{Morris_2018} showed that this modulation was consistent with the absence of dark spots, and with the presence bright spots with an effective temperature $\sim$5000K covering $\sim 8 \times 10^{-4}$ \% of the star's surface. Interestingly, \cite{Morris_2018} also noted a possible correlation between  the position of bright spots and flare events, casting doubt on the rotational origin of the 3.3d semi-periodic signal (see also \cite{Roettenbacher2017}). Nevertheless, recent MAROON-X spectroscopic observations of the transits of the TRAPPIST-1 planets revealed the system obliquity to be consistent with zero, while measuring a value of $2.1 \pm 0.3$ km s$^{-1}$ for the star's projected rotation velocity \citep{Brady_2023}. Assuming a null stellar inclination, this v$\sin{i}$ measurement translates into a rotation period of $2.9 \pm 0.5$ d, consistent with the candidate period  identified from photometry. 

Despite these numerous studies, and many others, the age, magnetospheric structure, and detailed photospheric abundances of the star remain poorly constrained. For instance, combining  constraints from measured kinematics, stellar density, spectral indices, rotation, activity, and metallicity, \citet{Burgasser_Mamajeck2017} estimated that the star is slightly metal-rich, it has an age of 7.6 $\pm$ 2.2 Gyr, and its 
magnetic activity is sufficient to influence its physical structure (see also \cite{Mullan2018}). Nevertheless, \citet{Gonzales2019} showed that some of the star's spectral features are indicative of a young, low-surface-gravity object, and proposed that this discrepancy could be due to the star's \hbindex{magnetic activity} or to the tidal influence of its planets.

Improving the characterization of the star TRAPPIST-1 through continued observations from the ground and from space will be required to better understand the formation and evolution of its planets. Notably, a drastic improvement of our understanding of the star's photospheric structure will be crucial for the thorough atmospheric characterization of its seven planets with JWST, as will be discussed below. 

Table ~\ref{tab:1} provides a summary of the parameters of the star TRAPPIST-1.

\begin{table}
\caption{Parameters of the star TRAPPIST-1.} 
\label{tab:1}       % Give a unique label
\begin{tabular}{p{3cm}p{3.4cm}p{1.2cm}p{4.5cm}}
\hline\noalign{\smallskip}
Parameter & Value + error & Unit & Reference \\
\noalign{\smallskip}\svhline\noalign{\smallskip}
Spectral type & M$8.0 \pm 0.5$ & & \cite{Gillon2016} \\
Mass $M_\ast$ & 0.0898 $\pm$ 0.0023  & $M_\odot$ & based on \cite{Mann2019} \\
Radius $R_\ast$ & 0.1192 $\pm$ 0.0013 & $R_\odot$ & \cite{Agol2021} \\
Density $\rho_\ast$ & 53.22 $\pm$0.53 & $\rho_\odot$ & \cite{Agol2021} \\
$\log_{10} (g_\ast$ [cm/s$^{-2}$])& $5.2936_{-0.0073}^{+0.0056}$ & dex & \cite{Agol2021} \\
Luminosity L$_\ast$ & $0.000552 \pm 0.000018$ & $L_{\odot}$ & \cite{Ducrot2020} \\
Distance & 12.467 $\pm$ 0.011 & pc & \cite{GaiaEDR3} \\
T$_{eff}$ & 2566 $\pm$ 26  & K & \cite{Agol2021} \\
$[Fe/H]$  & 0.04 $\pm$ 0.08 &  dex & \cite{Gillon2016} \\
v$\sin{i}$ & $2.1 \pm 0.3$  & km.s$^{-1}$ & \cite{Brady_2023} \\
P$_{rot}$ & 3.295 $\pm$ 0.004 & day & \cite{Vida2017}) \\
Age & $7.6 \pm 2.2$ & Gyr & \cite{Burgasser_Mamajeck2017} \\
$B$ & $600^{+200}_{-400}$ & G & \cite{Reiners2007} \\
$\log_{10}(L_{X}/L_{bol})$ & -3.52 $\pm$ 0.17 & dex & \cite{Weathley2017} \\
$\log_{10}(L_{XUV}/L_{bol})$ & -3.58 to - 3.44 & dex & \cite{Bourrier2017b} \\
$\log_{10}(L_{Ly_\alpha}/L_{bol})$ & -4.15 to -4.05 & dex & \cite{Bourrier2017b} \\
$\log_{10}(L_{H_\alpha}/L_{bol})$ & -4.85 to -4.60 & dex & \cite{Gizis2000, Reiners2007, Barnes2014, Burgasser2015} \\
\noalign{\smallskip}\hline\noalign{\smallskip}
\end{tabular}
\end{table}

\section{The TRAPPIST-1 planets}

\runinhead{Dynamics and formation.}

The seven TRAPPIST-1 planets form a super compact system with orbital distances ranging from 0.011 to 0.062 au. They have extremely coplanar and close to circular orbits \citep{Luger2017b, Agol2021} trapped in a complex \hbindex{resonance chain}. Indeed, the outer four planets are in first-order mean-motion resonances with each adjacent planet, while the inner three planets are near higher order resonances (8:5 and 5:3). Furthermore, each triplet of adjacent planets lie in (or close to) a 3-body Laplace resonances \citep{Luger2017, Teyssandier2022}.This dynamical configuration is reminiscent of a small number of multi-planetary systems found by Kepler and TESS (e.g. Kepler-80, \cite{MacDonald2016}; TOI-178, \cite{Leleu2021}). 
It may be indicative of a formation of the planets by \hbindex{pebble accretion} or \hbindex{planetesimals accretion} of the planets much further out in the original protoplanetary disk, simultaneous to or followed by their inwards \hbindex{migration} and their capture into resonance \citep{Ormel2017, Papaloizou2018, Coleman2019, Schoonenberg, Lin2021, Ogihara2022, Huang2022, Childs2023}. 

\runinhead{Densities and compositions.} 

\begin{figure}
\includegraphics[scale=0.37]{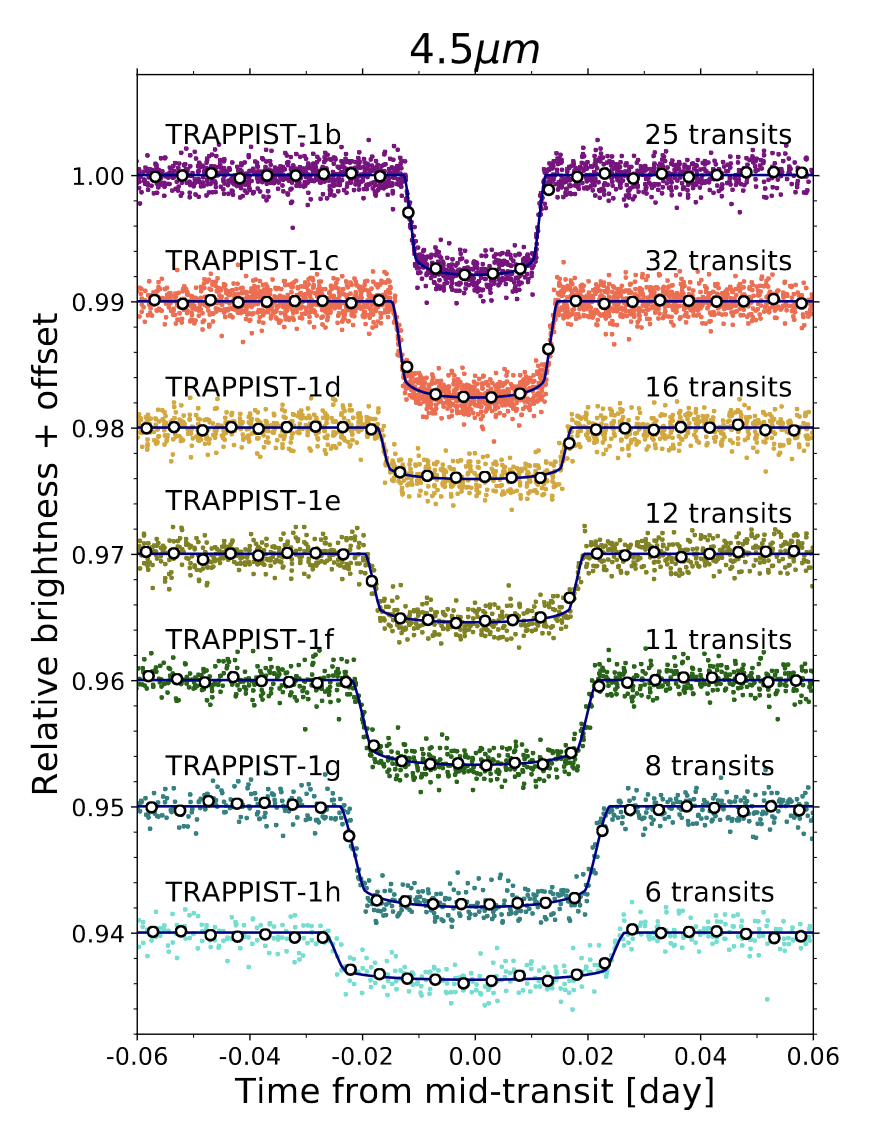}
\includegraphics[scale=0.26]{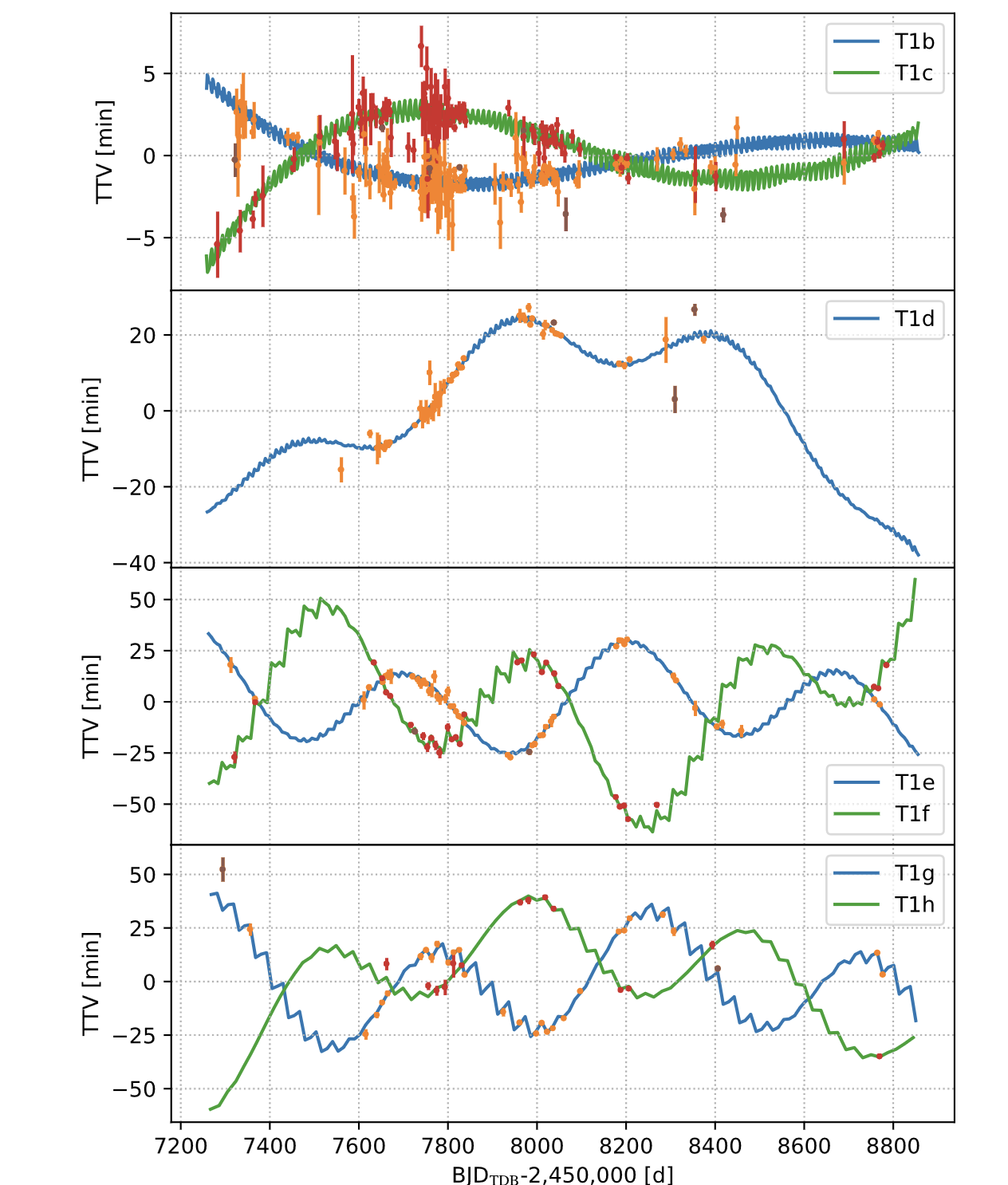}
\caption{$Left:$ combined transit light curves of the TRAPPIST-1 planets measured by {\it Spitzer} at 4.5 $\mu$m Source: open-source (arXiv) version of \cite{Ducrot2020}. $Right$: Transit Timing Variations (TTVs, orange/red error bars) of TRAPPIST-1 planets compared to the best-fit 8-body dynamical model (blue/green lines). Source: open-source (arXiv) version of \cite{Agol2021}.}
\label{fig:1}       % Give a unique label
\end{figure}

\begin{figure}
\includegraphics[scale=.35]{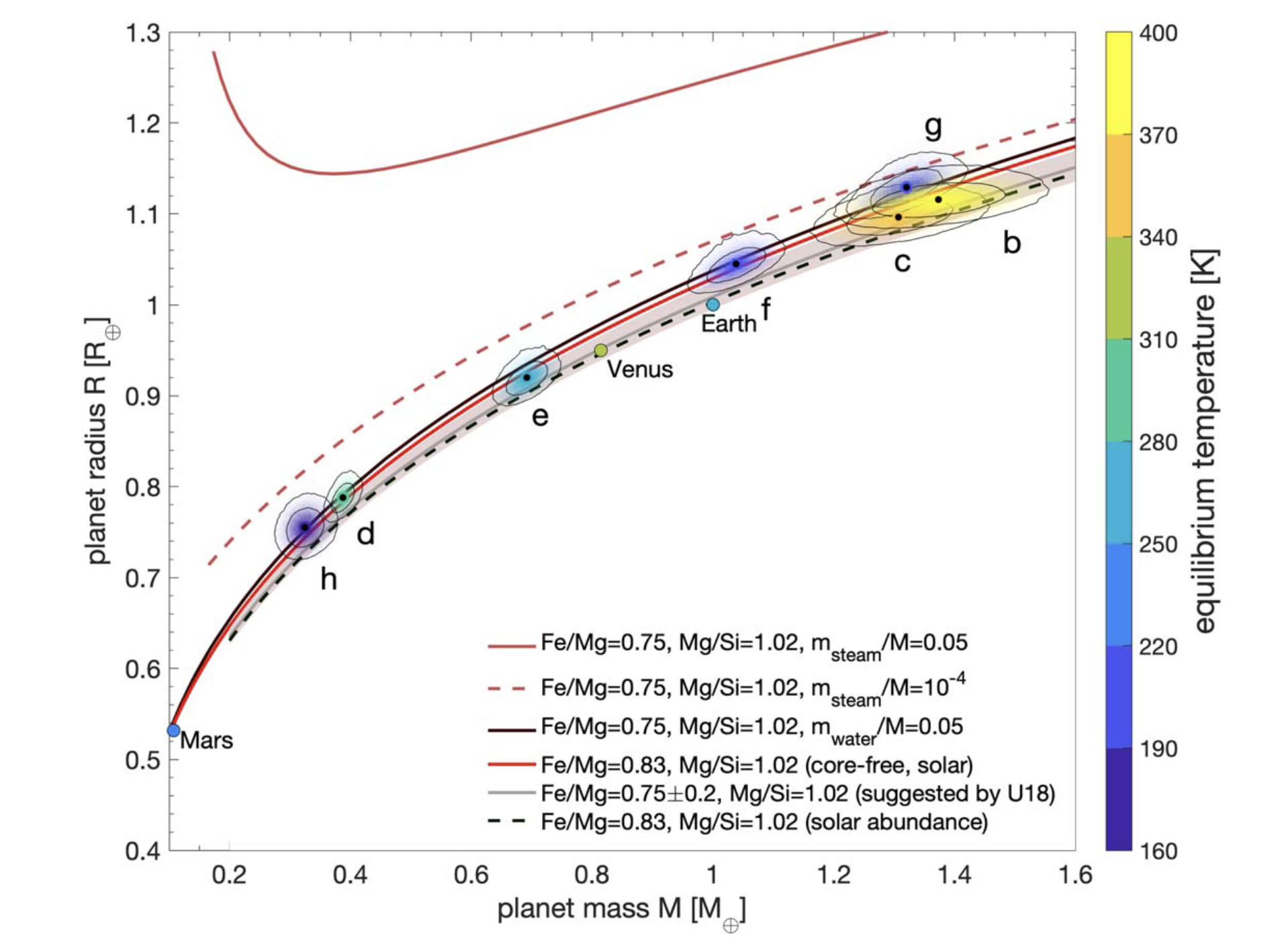}
\caption{Mass–radius relation for the seven TRAPPIST-1 planets as inferred by TTV and photodynamical analyses. Each planet’s posterior probability is colored by the equilibrium temperature (see color bar), with the intensity proportional to probability, while the 1$\sigma$ and 2$\sigma$ confidence levels from the posteriors are plotted with solid lines. Different theoretical mass–radius relations are overplotted as solid and dashed lines. Source: open-source (arXiv) version of \cite{Agol2021}} 
\label{fig:2}       % Give a unique label
\end{figure}

The observations of dozens of transits of the seven planets from the ground and from space led to very precise measurements of the planets' sizes (e.g. Fig. ~\ref{fig:1}, left). Furthermore, the resonant configuration of the system results in strong planet-planet interactions, leading to significant \hbindex{Transit Timing Variations} (TTV, \citealt{Agol2018}, Fig. ~\ref{fig:1}, right), that yields strong constraints on the planets' orbits and masses \citep{Gillon2017, Grimm2018, Agol2021}.  As a matter of fact, the TRAPPIST-1 planets are the rocky exoplanets with the most precise density measurements (see Fig. ~\ref{fig:2}). 

Initial TTV analyses yielded low densities for the
planets relative to an Earth-like composition (with the
exception of planet e) \citep{Grimm2018}. The most recent study of \citet{Agol2021} using a much extended transit timing data set led to drastically more precise (3-5 \%) mass measurements  (see Fig. ~\ref{fig:2} and Table  ~\ref{tab:2}). The resulting bulk densities for all seven planets are consistent with (1) a single rocky mass–radius relation which is depleted in iron relative to Earth, and otherwise Earth-like in composition, or with (2) an Earth-like composition enhanced in light elements, such as a surface water layer or a core-free structure with oxidized iron in the mantle \citep{Agol2021}.

\begin{figure}
\includegraphics[scale=.4]{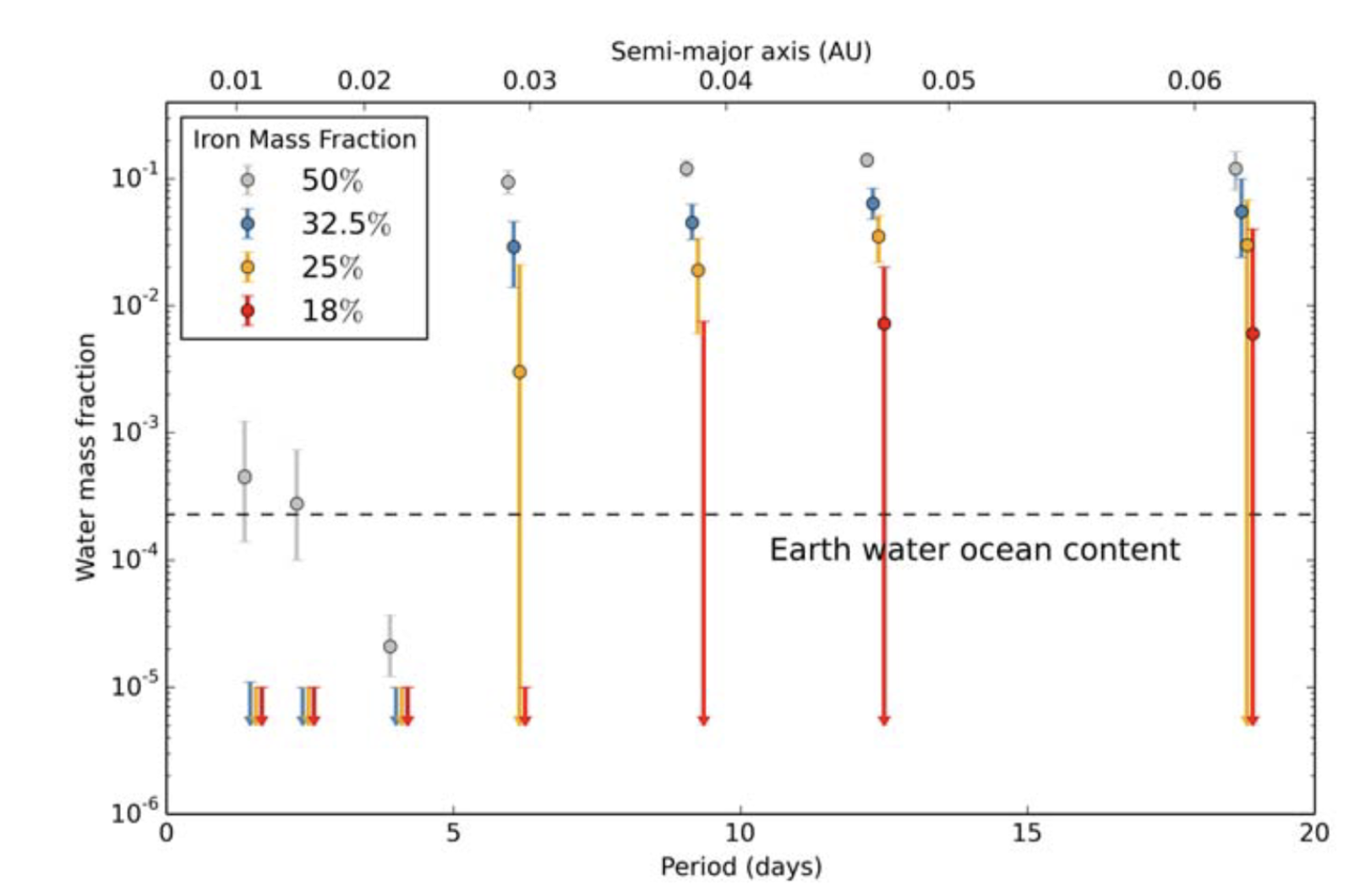}
\caption{Theoretical water content estimates (along with 1-$\sigma$ error bar) vs. planetary orbital periods in the TRAPPIST-1 system. Colors depict different core-mass fractions (CMFs) for the rocky interior.  Source: open-source (arXiv) version of \cite{Agol2021}} 
\label{fig:3}       % Give a unique label
\end{figure}

\cite{Agol2021} inferred that water-rich compositions are unlikely for the three inner planets, as they more irradiated than the runaway greenhouse irradiation limit \citep{Wolf2017, Turbet2018} for which all surface water should be vaporized, forming a thick H$_2$O-dominated steam atmosphere \citep{Turbet2019, Turbet2020} that should result in larger planetary radii than those measured. Assuming a core-mass-fraction similar to the Earth (32.5\%) or higher, \cite{Agol2021} also noted a possible increase of the inferred volatile content of the planets as a function of the orbital period (Fig. ~\ref{fig:3}) that could reflect that the outer planets formed beyond the water
condensation line at $\sim$0.025 au \citep{Unterborn2018} and could have experienced less dramatic water loss
(through atmospheric escape) thanks to their lower irradiations \citep{Bolmont2017}. This possibility is in line with the recent analysis of the formation of the TRAPPIST-1 system of \cite{Childs2023} that could reproduce the observed bulk densities of the two inner planets assuming that are nearly totally dessicated, while some significant volatile content was required for the five outer planets. 

\begin{table}
%\scriptsize
%
% Follow this input for your own table layout
%
\caption{Parameters of the TRAPPIST-1 planets. Source: \cite{Ducrot2020, Agol2021}} 
\label{tab:2}       % Give a unique label
\begin{tabular}{p{1.4cm}p{2cm}p{2cm}p{2cm}p{2cm}}
\noalign{\smallskip}\svhline\noalign{\smallskip}
Planet & b & c & d & e  \\
\noalign{\smallskip}\svhline\noalign{\smallskip}
$M$ $(M_\oplus)$ & $1.374 \pm 0.069$ & $1.308 \pm 0.056$ & $0.388 \pm 0.012$ & $0.692 \pm 0.022$  \\
$R$ $(R_\oplus)$ & $1.116 \pm 0.014$ & $1.097 \pm 0.014$ & $0.788 \pm 0.011$ & $0.920 \pm 0.013$ \\
$\rho$ $(\rho_\oplus)$ & $0.987 \pm 0.050$ & $0.991 \pm 0.043$ & $0.792 \pm 0.030$ & $0.889 \pm 0.033$ \\
$g$ $(g_\oplus)$ &  $1.102 \pm 0.052$ & $1.086 \pm 0.043$ & $0.624 \pm 0.019$ & $0.817 \pm 0.024$ \\
$S$ $(S_\oplus)$ & $4.15 \pm 0.16$  & $2.214 \pm 0.086$ & $1.115 \pm 0.043$&  $0.646 \pm 0.025$ \\
P (d) & $1.51088432$ & $2.42179346$ & $4.0497804$ & $6.0995648$ \\
 & $\pm 0.00000015$& $\pm 0.00000023$ & $\pm 0.0000027$ & $\pm 0.0000018$ \\
$a$ ($10^{-2}$ au) &  $1.154 \pm 0.010$ &$1.580 \pm 0.013$ & $2.227 \pm 0.019$ & $2.925 \pm 0.025$ \\ 
$i$ (deg) & $89.73 \pm 0.17$ & $89.78 \pm 0.12$ & $89.896 \pm 0.077$ & $89.793 \pm 0.048$ \\ 
\noalign{\smallskip}\svhline\noalign{\smallskip}
\noalign{\smallskip}\svhline\noalign{\smallskip}
Planet & f & g & h & \\
\noalign{\smallskip}\svhline\noalign{\smallskip}
$M$ $(M_\oplus)$  & $1.039 \pm 0.031$ & $1.321 \pm 0.038$ & $0.326 \pm 0.020$ & \\
$R$ $(R_\oplus)$ & $1.045 \pm 0.013$& $1.129 \pm 0.015$& $0.755 \pm 0.014$ & \\
$\rho$ $(\rho_\oplus)$ & $0.911 \pm 0.029$& $0.917 \pm 0.029$ & $0.755 \pm 0.059$ & \\
$g$ $(g_\oplus)$ &  $0.951 \pm 0.024$& $1.035 \pm 0.026$ & $0.570 \pm 0.038$  & \\
$S$ $(S_\oplus)$ &   $0.373 \pm 0.015$ & $0.252 \pm 0.010$ & $0.144 \pm 0.006$ & \\
P (d) & 9.2065940 & 12.3535557 & 18.767275 & \\
& $\pm 0.0000021$ & $\pm 0.0000034$ & $\pm 0.000019$ & \\ 
$a$ ($10^{-2}$ au) & $3.849 \pm 0.033$ & $4.683 \pm 0.040$ & $6.189 \pm 0.053$ &  \\ 
$i$ (deg) & $89.740 \pm 0.019$ & $89.742 \pm 0.012$& $89.805 \pm 0.013$ & \\ 

\noalign{\smallskip}\hline\noalign{\smallskip}
\end{tabular}
\end{table}
%

\runinhead{Atmospheres?} 
The frequency of terrestrial planets orbiting in the \hbindex{habitable zone} of mid-to-late-type M-dwarfs is still poorly constrained, especially for late-type M-dwarfs, but given their high occurrence rate of terrestrial planets on shorter orbital periods \citep{Ment2023} and their tendency to host multiple compact systems of planets, it is likely that a significant fraction of mid-to-late-type M-dwarfs harbor at least one planet in their habitable zone. This hypothesis is consistent with the discovery of a terrestrial planet orbiting in the habitable zone of our nearest neighbor star, the M6-type dwarf \hbindex{Proxima Centauri} \citep{Anglada2016}. Considering that most stars in the solar neighborhood are mid-to-late-type M-dwarfs \citep{Henry2006}, one can thus hypothesize that the majority of potentially habitable planets in the Milky Way orbit around such tiny stars. Nevertheless, the actual habitability of these planets is far from certain \citep[e.g.][]{Shields2016}. The most serious fear in this regard concerns the capacity of these planets to maintain a significant atmosphere. Indeed, most mid-to-late-type M-dwarfs are magnetically active, resulting in XUV irradiations and stellar winds strong enough, in theory, to significantly \hbindex{erode} the secondary atmosphere of a rocky planets orbiting near or within the circumstellar habitable zone within less than a Gyr (e.g. \citealt{Airapetian2017, Garcia-Sage2017}). Furthermore, their pre-main-sequence phase can extend up to 1 Gyr during which planets later on temperate orbits could undergo a runaway greenhouse phase able to dessicate them completely or, at least, to compromise their potential habitability \cite{Luger2015}. Nevertheless, current atmospheric erosion models have large theoretical uncertainties, and they predict that early Earth suffered massive volatile losses for which no clear imprint is found in the present volatile inventory (see discussion in \citealt{Ribas2016}). Furthermore, rocky planets around low-mass M-dwarfs could have an initial volatile content orders of magnitude larger than the Earth whose \hbindex{outgassing} could efficiently counterbalance the erosion and maintain a  significant secondary atmosphere in a steady state (e.g. \citealt{Bolmont2017}).

In this context, the TRAPPIST-1 system represents a unique laboratory to assess the capacity of Earth-sized terrestrial planets in short orbits around a late-type M-dwarf to maintain a significant atmosphere, and to develop habitable surface conditions for those orbiting in the circumstellar habitable zone. Indeed, its planets have an irradiation range similar to the inner solar system and encompassing the inner and outer limits of its circumstellar habitable zone, with planet b and h receiving from their star about 4.2 and 0.15 times the energy received by the Earth from the Sun per second, respectively. Detecting an atmosphere around any of these 7 planets and measuring its  composition would be of fundamental importance to constrain our atmospheric evolution and escape models, and, more broadly, to determine if low-mass M-dwarfs, the larger reservoir of terrestrial planets in the Universe, could truly host habitable worlds. 

A first search for atmospheres around the TRAPPIST-1 planets 
was performed with HST/WFC3 (Programs 15400, 14873, 15304). Due to the precision and spectral limitations of HST/WFC3, this search
was limited to primordial cloud-free \hbindex{hydrogen-dominated atmospheres}. 
The observations ruled out the presence of such clear
extended atmospheres for all TRAPPIST-1 planets \citep{deWit2016, dewit2018, Wakeford2019, Gressier2022, Garcia2022}.
In parallel, HST’s UV capabilities were leveraged to search for the presence of hydrogen exospheres around the planets (program 14900 \& 15304). Here again, the observations obtained for all seven planets failed  to reach a detection, while placing upper limits on their current atmospheric escape rates \citep{Bourrier2017b, Bourier2017a}.

Many studies showed that JWST should have the instrumental potential
to detect \hbindex{secondary compact atmospheres} for all TRAPPIST-1 planets
\citep[e.g.][]{Barstow2016, Morley2017, Batalha2018, Krissansen2018, Fauchez2019, Wunderlich2019, Lustig-Yaeger2019}. These studies
converged on the following picture:  significant secondary atmospheres could be detected by JWST in transmission –and in emission
for the two inner planets– assuming the observation
of a number of \hbindex{transits/occultations} ranging from less
than ten to more than one hundred, depending on the
atmospheric properties of the planets (composition,
cloud coverage). These studies motivated several JWST Cycle 1 programs
to target TRAPPIST-1 planets: GTO programs  1177 (5 occultations of planet b with MIRI at 15 $\mu$m), 1201 
(2 transits of planet d with NIRSPEC, 5 transits of planet f 
with NIRISS), 1279 (5 occultations of planet b with MIRI at 12.8 $\mu$m), 1331 (4 transits of planet e with NIRSPEC), and GO programs 1981 (3 transits of planet h with NIRSPEC), 2304 (4 occultations of planet c with MIRI), 2420 (4 transits of planet c with NIRSPEC), 2589 (2 transits of planet b and of planet c with NIRISS, 3 transits of planet g with NIRSPEC). In Cycle 2, program 3077 will use MIRI to measure the combined \hbindex{phase curve} of planets b and c at 15 $\mu$m. 

At the time of writing, only three JWST results have been published for TRAPPIST-1. The two first ones were obtained by \hbindex{emission occultation photometry} with MIRI. By observing five occultations of planet b at 15 $\mu$m, \citet{Greene2023Natur} firmly detected the thermal emission of the planet and measured a high dayside brightness temperature of 503 $\pm$ 27K suggestive of a poor distribution of heat to the night side, and consistent with the emission of a null-albedo bare-rock scenario and with some low-density atmosphere scenarios \citep{Ih2023}. On their side, \cite{Zieba2023} observed four occultations of planet c, again at 15 $\mu$m. Their measured dayside brightness temperature of 380 $\pm$ 31K disfavours a dense Venus-like CO2-rich atmosphere or a null albedo bare rock scenario, and leaves room for some alternative atmospheric (CO2-poor, low density) and airless (high-albedo surface) scenarios. 

\begin{figure}
\includegraphics[scale=.5]{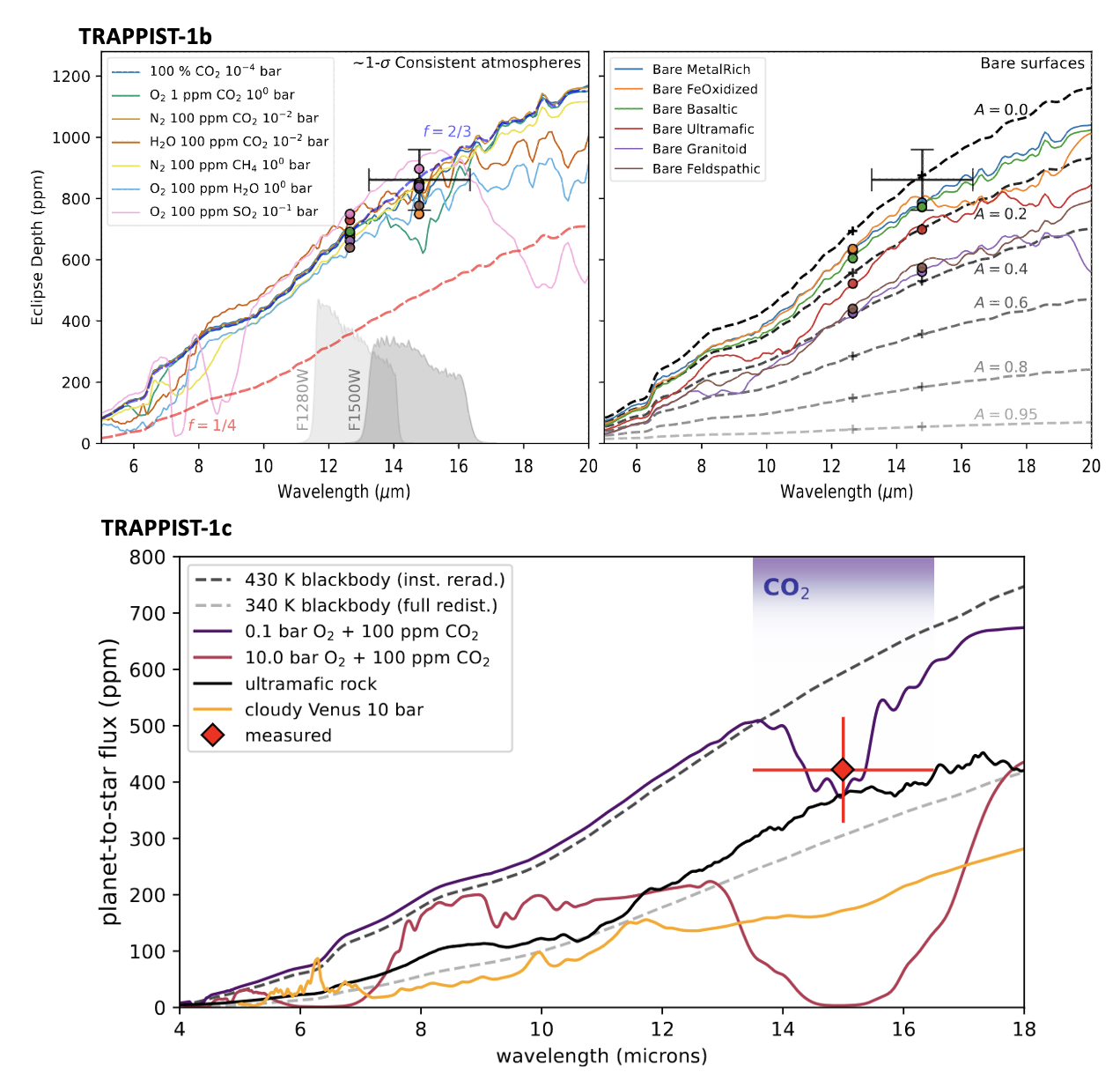}
\caption{Occultation depths measured at 15 $\mu$m for TRAPPIST-1\,b by GTO program 1177 \citep{Greene2023Natur} (top) and for TRAPPIST-1\,c by GO program 2304 \citep{Zieba2023} (bottom). Vertical error bars are the 1-$\sigma$ uncertainties. Horizontal error bars represent the extent of the MIRI F1500W filter.  For both planets, atmospheric and surface models are compared to the measurements. In addition to some low-density atmosphere scenarios, planet b's measurement is consistent with a null-albedo bare surface, while for planet c an airless scenario  would favor a Bond albedo $>$ 0.3.  Source: open-source (arXiv) versions of \cite{Ih2023, Zieba2023}.}
\label{fig:4}       % Give a unique label
\end{figure}

It is not yet possible to reach a decisive conclusion on the presence/absence of an atmosphere from these single photometric measurements, but they already reveal a particularly interesting first picture: planet b and c seem to be different, with a higher posterior probability for the presence of an atmosphere or for a larger surface Bond Albedo for the outer one (Fig. \ref{fig:4}). Unfortunately, JWST is `too small' to obtain a precise emission spectrum of the two planets by occultation spectroscopy. The best it could do is to obtain such photometric measurements in two or three more bandpasses. This is exactly what GTO program 1279 will provide: measuring the brightness temperature of planet b's dayside at 12.8 $\mu$m to obtain an extra-constraint on the presence of a scarce atmosphere or on the planet's surface composition. Cycle 2 Program 3077 should lead to a more robust inference on the presence of an atmosphere around planet b and c by measuring their combined thermal phase curve at 15 $\mu$m that will immediately inform on the efficiency of the transport of heat to the night side of both planets.

\begin{figure}
\includegraphics[scale=.34]{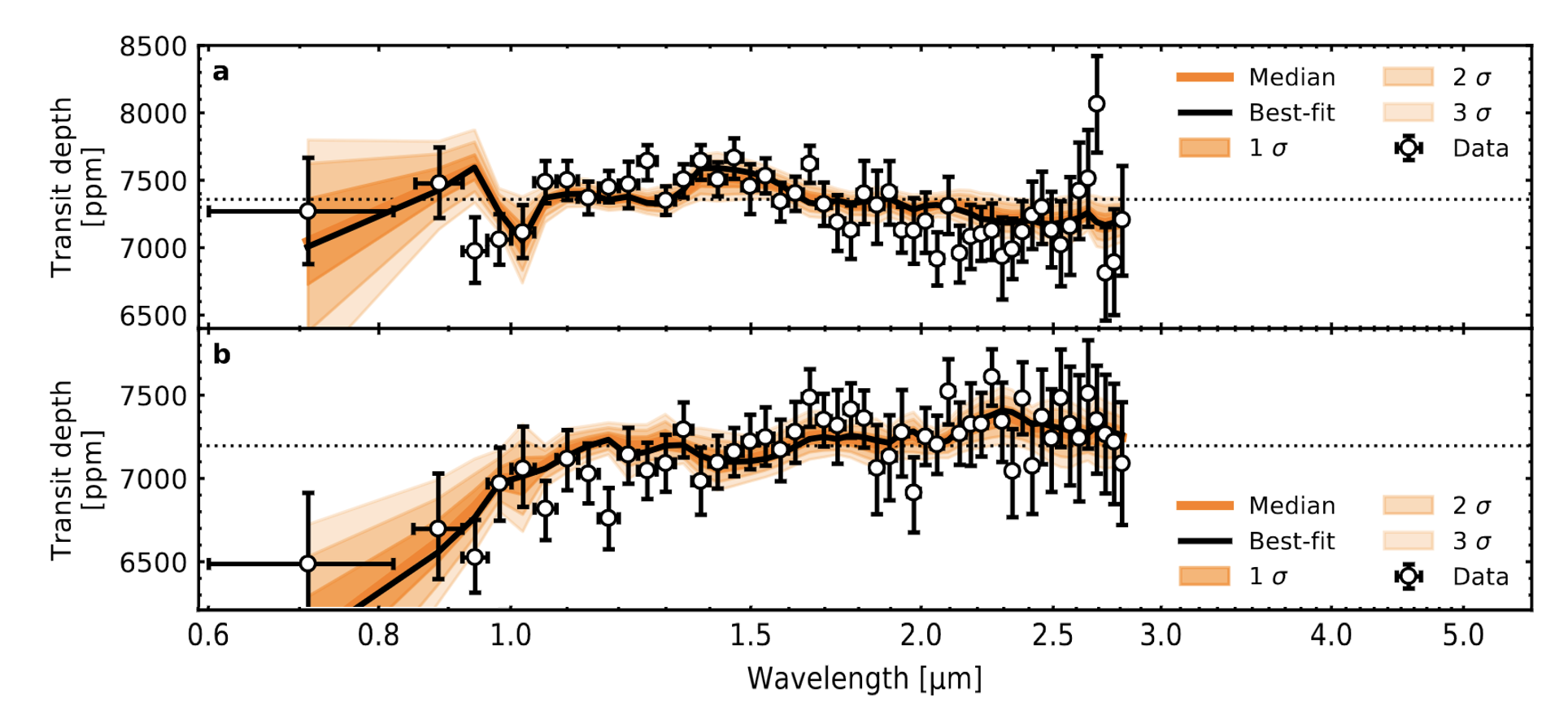}
\caption{Transit transmission spectra of TRAPPIST-1\,b as measured by JWST NIRISS/SOSS within GO program 2589. Each panel corresponds to a specific visit. Vertical error bars are the 1-$\sigma$ uncertainties. Horizontal error bars represent the extent of each spectral
bin. For both visits, the transmission spectrum is compared to the best-fit/median stellar contamination model (black/orange curves) and uncertainties (shaded regions). Source: \cite{Lim2023}} 
\label{fig:5}       % Give a unique label
\end{figure}

On the \hbindex{transit transmission spectroscopy} side, a first result was published by \cite{Lim2023} for planet b. Two transits of the planet were observed with NIRISS (Program GO 2589) in the SOSS mode covering the 0.6 - 2.8 $\mu$m spectral range at a resolving power of $\sim$700. The two obtained transmission spectra (Fig. \ref{fig:5}) are dominated by \hbindex{stellar contamination} from unocculted photospheric heterogeneities. Indeed, when a planet transits a star with surface heterogeneities (spots or faculae), the flux from the transited chord may not be representative of the
flux from the full stellar disk, resulting in features in the transmission spectrum able to bias the planetary atmosphere retrieval \citep[e.g.][]{Sing2011, McCullough2014, Rackham2018, Rackham2023}. The significant difference between the two visits' spectra visible in Fig. \ref{fig:5} are well explained by the contamination of an unocculted spot in the first visit and the contamination of an unocculted faculae in the second visit. These results strongly support previous evidence that \hbindex{stellar contamination} will be a critical consideration in the study of TRAPPIST-1 planets by transit transmission spectroscopy \citep[e.g.][]{Zhang2018, Wakeford2019, Ducrot2020, Garcia2022}. Indeed, the study found that the  uncertainties associated with the lack of stellar photospheric model fidelity are one order of magnitude above the observation precision, preventing any conclusion on the presence/absence of a secondary atmosphere to be drawn. Only a primary hydrogen-dominated atmosphere could be discarded, in agreement with previous HST results. 

As more JWST transmission spectroscopy results will appear, the amplitude of the challenge posed by stellar contamination will get clearer. Nevertheless, these first results already outline the critical need for additional theoretical and observational work to constrain the photospheric structure of the star \citep{Rackham2023} to make possible isolating atmospheric signatures from stellar contamination. In this context, the TRAPPIST-1 system presents another advantage: its seven planets cover a relatively large range of impact parameters (from $\sim$0 to 0.4), and the global analysis of their transmission spectra should thus bring key constraints on the global photospheric structure of the star. If the absence of a significant atmosphere around the two inner planets is confirmed with more observations, they could be used as `probes' to map the stellar \hbindex{photospheric structure}. In practice, it would require to group observations of their transits and of transits of the outer planets, and to use  their transit spectra to constrain the structure of the stellar photosphere at the time of the observations. The atmospheric study of TRAPPIST-1 planets with JWST should thus be more challenging than previously thought, but in return it should improve our knowledge of the magnetic and photospheric structure of ultracool dwarf stars, one of the most frequent kind of stars of the Universe.

%\section{Conclusions} 

%\begin{svgraybox}
%The TRAPPIST-1  system is composed of seven Earth-sized rocky planets in short orbits around a Jupiter-sized ultracool dwarf star 12 parsec away. These planets' irradiations range is similar to the one of the inner solar system, and three of them orbit within the circumstellar habitable zone . Thanks to the small size and to the infrared brightness of the host star, and to the system's compact resonant structure, these planets are particularly well-suited for a detailed characterization. An intense transit timing monitoring campaign resulted in unprecedented precisions on the planet's masses and densities and in strong constraints on their compositions. Transit transmission spectroscopy with HST discarded the presence of extended primary atmospheres around the seven planets. First results obtained with JWST favor low-density atmosphere or bare rock scenarios for the two inner planets. The detection of dense secondary atmospheres around the five outer planets could be achieved with JWST, but only if the critical problem of stellar contamination can be addressed with more theoretical and observational work.  \end{svgraybox}

\section{Cross-References}
\begin{itemize}
\item{Transit Photometry as an Exoplanet Discovery Method}
\item{Transit-Timing and Duration Variations for the Discovery and Characterization of Exoplanets}
\item{SPECULOOS Exoplanet Search and Its Prototype on TRAPPIST}
\item{The Impact of Stellar Activity on the Detection and Characterization of Exoplanets}
\item{Exoplanet Atmosphere Measurements from Transmission Spectroscopy and Other Planet Star Combined Light Observations}
\item{Proxima b: The Detection of the Earth-Type Planet Candidate Orbiting Our Closest Neighbor}
\item{The Habitable Zone: The Climatic Limits of Habitability}
\end{itemize}

\begin{acknowledgement}
The author is F.R.S.-FNRS Research Director. The author thanks Eric Agol for his thorough review of the manuscript.
\end{acknowledgement}

\bibliographystyle{spbasicHBexo}  %for bibtex
\bibliography{gillon.bib} %for bibtex-example

\end{document}